\documentclass[letterpaper,twocolumn,fleqn]{article} 

\usepackage{cite}
\usepackage{amsmath,amssymb,amsfonts}
\usepackage{algorithmic}
\usepackage{graphicx}
\usepackage{textcomp}
\usepackage{xcolor}
\usepackage{todonotes}
\usepackage[utf8]{inputenc}
\usepackage{ist}
\usepackage{subcaption}
\usepackage[justification=centering]{caption} %center table caption
\title{Interactive Multi-User 3D Visual Analytics in Augmented \\
Reality}

\author{Wanze Xie\textsuperscript{2}, Yining Liang\textsuperscript{2}, Janet Johnson\textsuperscript{2}, Andrea Mower\textsuperscript{1}, Samuel Burns\textsuperscript{1}, Colleen Chelini\textsuperscript{1}, Paul D'Alessandro\textsuperscript{1}, Nadir Weibel\textsuperscript{2}, J\"urgen P. Schulze\textsuperscript{2} \\
\textsuperscript{1}BodyLogical team \\
\textsuperscript{2}University of California San Diego, La Jolla, CA, USA}

\date{} % date has an empty field.

\begin{document} 

\maketitle 

\thispagestyle{empty} % prevents the first page to be numbered

%%%%%%%%%%%%%%%%%%%%%%%%%%%%%%%%%%
% Abstract
%%%%%%%%%%%%%%%%%%%%%%%%%%%%%%%%%%
\begin{abstract}
This publication reports on a research project in which we set out to explore the advantages and disadvantages augmented reality (AR) technology has for visual data analytics. We developed a prototype of an AR data analytics application, which provides users with an interactive 3D interface, hand gesture-based controls and multi-user support for a shared experience, enabling multiple people to collaboratively visualize, analyze and manipulate data with high dimensional features in 3D space. Our software prototype, called DataCube, runs on the Microsoft HoloLens - one of the first true stand-alone AR headsets, through which users can see computer-generated images overlaid onto real-world objects in the user's physical environment. Using hand gestures, the users can select menu options, control the 3D data visualization with various filtering and visualization functions, and freely arrange the various menus and virtual displays in their environment. The shared multi-user experience allows all participating users to see and interact with the virtual environment, changes one user makes will become visible to the other users instantly. As users engage together they are not restricted from observing the physical world simultaneously and therefore they can also see non-verbal cues such as gesturing or facial reactions of other users in the physical environment. The main objective of this research project was to find out if AR interfaces and collaborative analysis can provide an effective solution for data analysis tasks, and our experience with our prototype system confirms this.
\end{abstract}

\section{Introduction} \label{sec:intro}

The goal of this project was to design an augmented reality data visualization tool for population health management, which uses the biomedical simulation tool Bodylogical \cite{Bodylogical} to predict the population's future health by scientifically simulating each individual's health based on dozens of biomarkers such as blood pressure, cholesterol, glucose levels, etc. In this article we will report on how we created our augmented reality prototype system, along with the challenges we encountered and the insights we gained along the way.

Bodylogical is a simulation tool, which for each member of the simulated population, creates a digital twin with simulated values for many of the human biomarkers and bodily functions. By giving lifestyle choices, such as the amount of calories taken in, the amount of exercise or the amount of sleep, the simulation tool can predict the future health of each simulated individual, and thus the entire population.

%%%%%%%%%%%%%%%%%%%%%%%%%%%%%%%%%%
% Related Work
%%%%%%%%%%%%%%%%%%%%%%%%%%%%%%%%%%

\section{Related Work}

There are two main areas of prior work our project is built upon. One is real-time, multi-user augmented reality data visualization, the other is visual analytics.

The release of the Microsoft HoloLens \cite{HoloLens} in 2016 defined a point in the history of augmented reality from which on it was possible to build augmented reality applications which could run directly on a headset, did not require an external computer, and they allowed the programmer to focus on the user facing part of the software application because the HoloLens had built-in inside-out tracking which was very good, as well as a built-in spatial user interface. It even has its own finger tracking parser, which allows for simple but effective interaction with AR applications by interpreting the user's head direction as a cursor while finger gestures trigger events.

Since the release of the HoloLens, researchers and application developers around the world have developed numerous AR applications and wrote academic publications on them. Standout areas of particular interest are medical visualization, molecular science, architecture and telecommunication. The AR application OnSight which allows scientists to virtually walk on planet Mars, a collaboration between Microsoft and JPL, even received NASA's 2018 Software of the Year Award \cite{OnSight}. It allows multiple users, each with a HoloLens, to explore the surface of Mars. 

Trestioreanu \cite{Trestioreanu2018} created a collaborative medical data visualization system for the HoloLens which can visualize CT data sets in 3D, but it is not designed to do general data visualization. And it requires an external computer for the actual rendering task, as opposed to rendering directly on the HoloLens headset.

Visual data analytics in 3D is not a novel concept. It doesn't require AR but can just as soon be done on a regular monitor or in virtual reality (VR). Millais et al. \cite{Millais2018} compare 3D data visualization in VR to traditional 2D visualization and found that 3D visualization increases the accuracy and depth of insights compared to 2D. To us, the collaborative aspect of AR data visualization is also very important, because it allows groups of experts to brainstorm together while viewing the data in front of them.

%%%%%%%%%%%%%%%%%%%%%%%%%%%%%%%%%%
% Implementations
%%%%%%%%%%%%%%%%%%%%%%%%

\section{Implementation}

On the pathway to our final HoloLens application, we developed two separate prototype applications. 

\subsection{Prototype 1: Bar Charts}

\begin{figure}[htb]
    \centering
    \includegraphics[width=\linewidth]{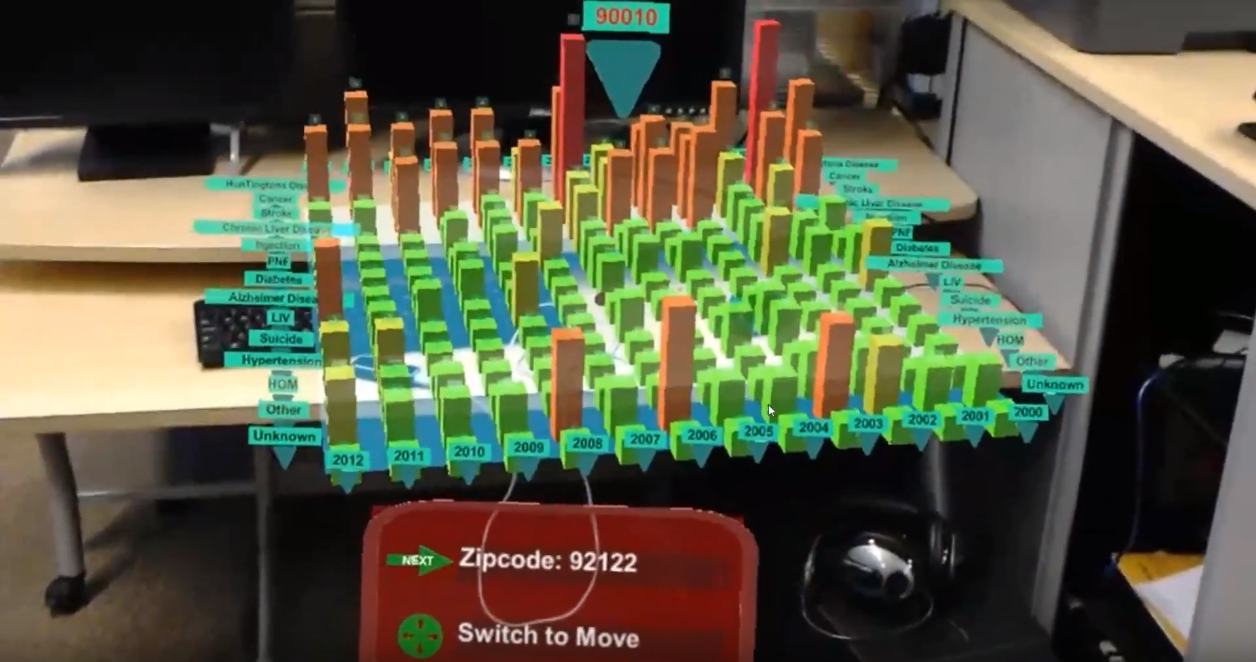}
    \caption{A 3D bar chart in augmented reality on the HoloLens.}
    \label{fig:3d-barchart}
\end{figure}

The first prototype is an AR application which renders a 3D bar chart in the AR environment, and places it on a physical table in the room that the user chooses within the application. The purpose of this is that multiple users can be standing around the visualization and independently view it from their respective viewing directions, just as if there was a physical model on the table between them. This early version of the visualization tool already had basic features such as loading data files, selecting data points, and the selection of viewing and filtering parameters of the 3D bar chart. Figure \ref{fig:3d-barchart} shows an image of what this application looks like from the user's point of view, as seen through the HoloLens.

This approach has some advantages of identifying standout data groups by looking at taller bars, and also provides a quick reference to the data group of a certain year for a specific region, selected by zip code. In addition to bar height, we use a color gradient from cool colors (blue, green) to warm colors (yellow, red) to emphasize the differences between bar heights, because they are otherwise sometimes difficult to discern due to occlusion effects when the user looks at the data from a low viewpoint, or lack of discernible bar heights when the user looks from the top down. But it turns out that there are many limitations to this kind of visualization, primarily that it does not provide much information in addition to what a typical Excel chart on a 2D display would show.

\begin{figure}[htb]
    \centering
    \includegraphics[width=79mm]{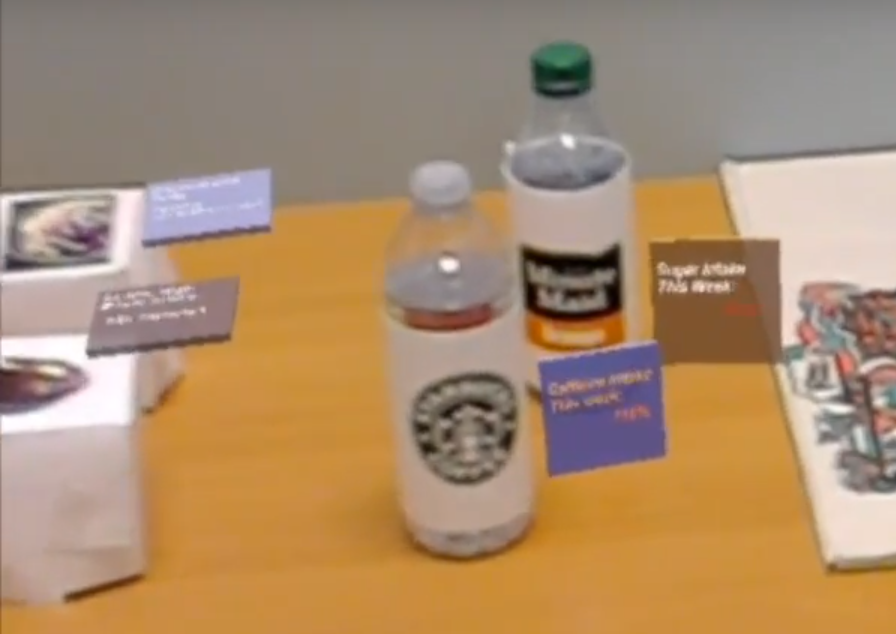}
    \caption{Embedded Visualization based on image recognition.}
    \label{fig:embedded-visualization}
\end{figure}

Within this prototype, we also explore ways to make better use of the environment provided by the augmented reality system. Instead of laying out all the data on a flat surface, we followed Bach et al.'s concept of “Embedded Visualization” \cite{bach2017drawing} (see Figure \ref{fig:embedded-visualization}). As a result, our AR application targets individual users' health data, rather than aggregate health information averaged over larger amounts of people. Our visualization tool provides Embedded Visualization based on image and object recognition to show data relevant to the health of a user.

\subsection{Prototype 2: DataCube}

\begin{figure}[htb]
    \centering
    \includegraphics[width=\linewidth]{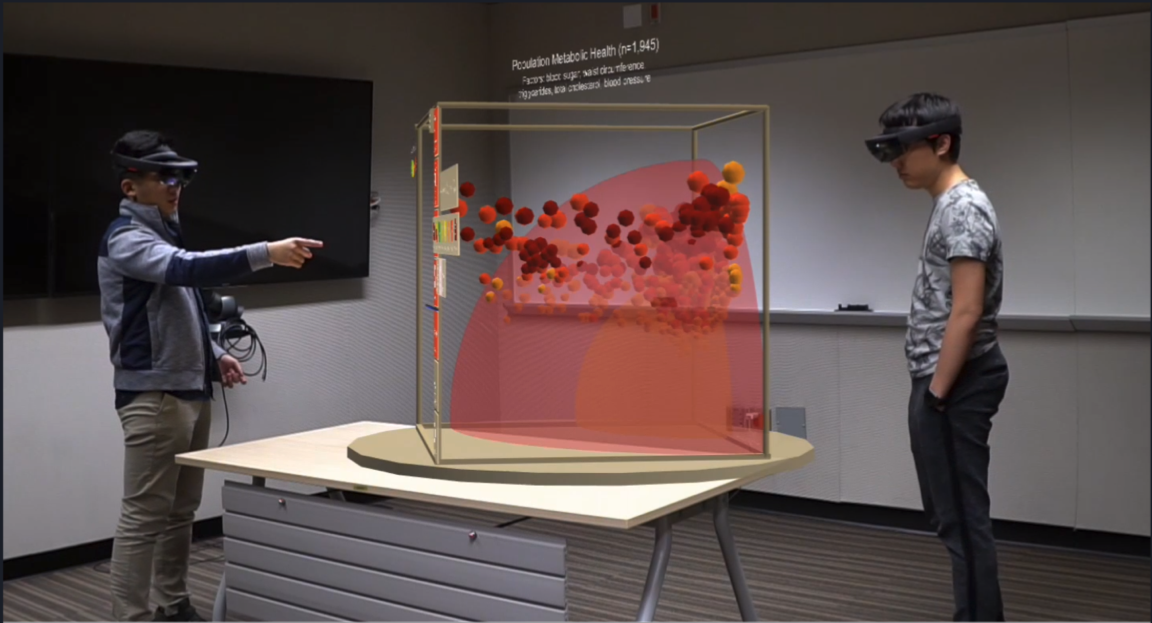}
    \caption{Our AR data visualization tool DataCube with two users (one looking at the other).}
    \label{fig:datacube}
\end{figure}

Our second prototype is a group data visualization tool, and it is much more sophisticated than our initial 3D bar chart application. We used the lessons we learned from the first prototype, such as positioning 3D data sets on a physical table in the room, or menu items that are attached to the data visualization object to build a multi-user application which allows an analyst to analyze data in truly novel ways. The core component of this design is a multi-dimensional scatter plot, displayed inside of our DataCube, which can display six dimensions of data by making use of the three spatial coordinate axes, as well as color, sphere size and trace lines. Figure~\ref{fig:datacube} shows what our DataCube application looks like when two people are in a collaborative session.

\begin{figure}[htb]
    \centering
    \includegraphics[width=\linewidth]{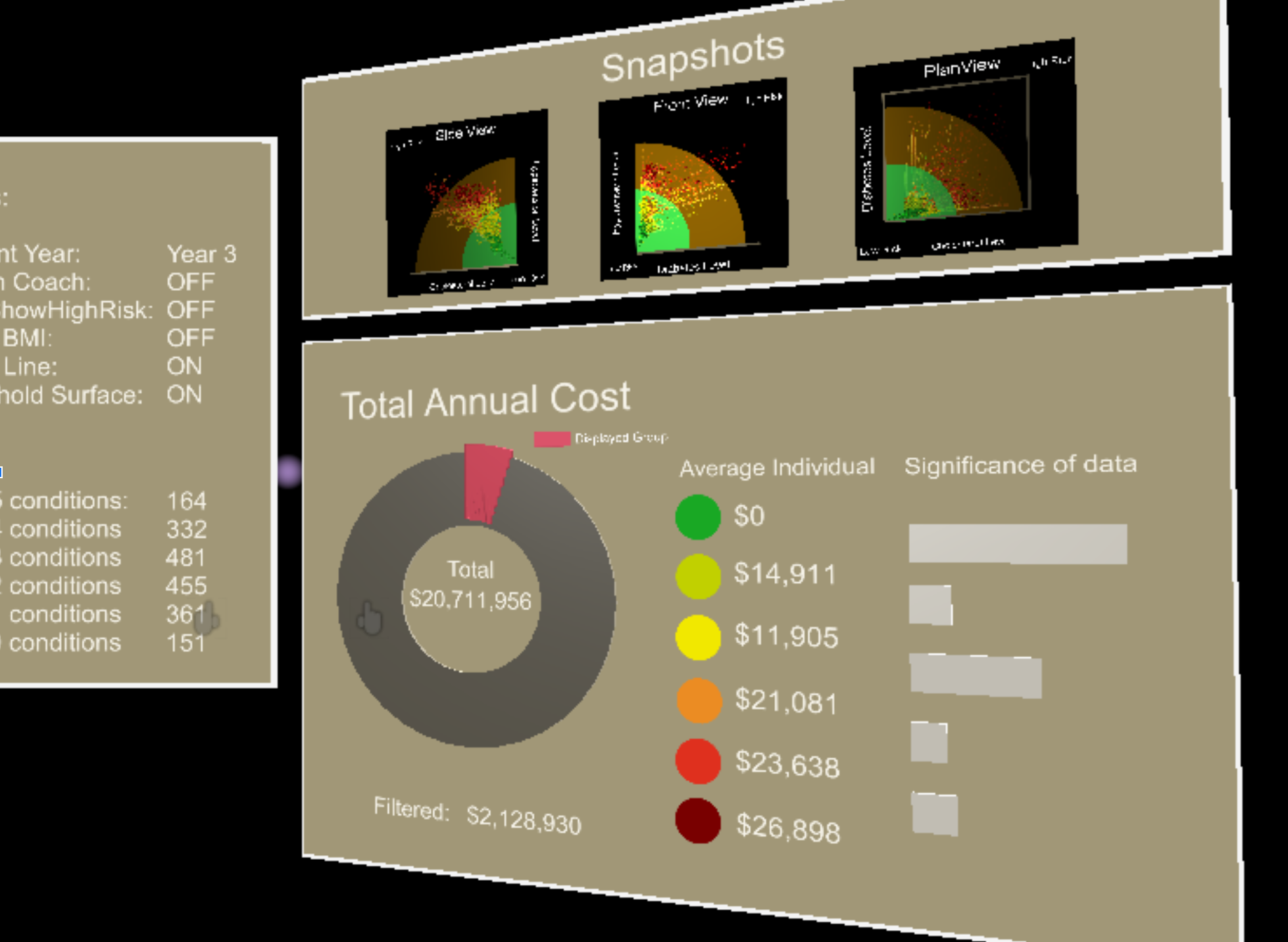}
    \caption{The analysis wall is placed on one of the walls of the physical room.}
    \label{fig:wall}
\end{figure}

Another component of our visualization tool is the analysis wall, see Figure~\ref{fig:wall}. It is placed on one of the walls of the room the users are in, so that anyone can walk up to it to take a closer look and analyze subsets of the data with 2D diagrams and statistical information. The user can also choose to create a snapshot of what they are viewing inside the DataCube. The snapshots are 2D scatter plots based on the face of the DataCube the user looks at, allowing the user to isolate crucial aspects of the data or the visualization.

\subsubsection{File System}

The DataCube application uses comma-separated value (CSV) files to store the population health data. These files are stored in a OneDrive data directory in the cloud, linked to the HoloLens's Microsoft account. The authentication process is secure and managed by the HoloLens at the operating system level. 

To import a data file, the user clicks on the import button on the virtual console and the HoloLens will open up the "Load from OneDrive" menu for the user to select a file. The data file needs to be uploaded by the user to their OneDrive and is required to follow our custom CSV file format to work with our data importer. During or at the end of an analysis session, the user can take snapshots of the data and export them to a file for further investigation by clicking the "Export" button in the main DataCube menu, and the HoloLens will open up the "Send as email" window and attach the current data visualization converted to a file as an attachment. In this window, the user can enter the recipient's email address and click send. All of this happens within the HoloLens application, interactions are done via the standard HoloLens interface of pointing by head orientation and finger pinching to trigger actions. 

\subsubsection{Shared Multi-User Experience}

\begin{figure}[htb]
    \centering
    \includegraphics[width=\linewidth]{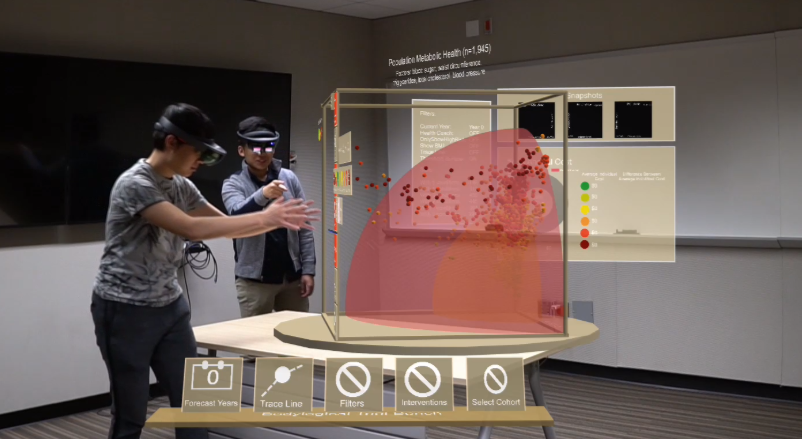}
    \caption{Third person view of two users using the application in a collaborative session.}
    \label{fig:datacube-2users}
\end{figure}

Another feature of our DataCube prototype is the shared multi-user experience. Up to five people, each wearing a HoloLens, can connect into the same session and analyze, discuss and manipulate the visualization together in the same place at the same time (see Figure \ref{fig:datacube-2users}).

To set up the shared experience, we implemented a socket-based communication protocol within the local area network (LAN) based on the MixedReality Toolkit provided by Microsoft. The sharing mechanism is server-based. We set up the sharing server on a separate PC and require all participating HoloLens users to be connected to the same LAN as the PC. When the DataCube application is launched, each HoloLens will automatically search for the server PC's host name within the LAN and connect to it. Since all HoloLenses are required to be on the same wireless LAN, all users will need to be able to log in to the same wifi network, which ensures that all parties in the session are trusted users.

The sharing is also session-based. The first user defines the 3D coordinate system the participants will be sharing by sending its spatial coordinate system's common points of reference (anchors) to the server. When subsequent users join the session, they will automatically download the anchor information stored on the server and try to match the anchor points with their own spatial coordinate system. Eventually, each party in the same session will share the same coordinate space, provided that they are physically in the same room.

We keep track of user interactions and synchronize the interactions that are applied to the shared objects. This include the transformation matrices of movable objects and users, any change of states of the data analysis cube and the analysis wall, etc. The menu of the data cube does not have a shared transformation matrix, but we design it in a way so that regardless of the direction the user faces with respect to the cube, the main menu always faces to the user to make it as easy as possible to select menu items.

One limitation of the HoloLens shared experience is the number of participants in the same session due to the potential interference across multiple HoloLens devices. According to the official Microsoft HoloLens documentation \cite{holoLensdoc}, the recommended maximal number of HoloLenses in the same shared experience session is six. We tested up to five users in the same session and were able to achieve a stable connection between all users and the server for the duration of typical user sessions, which can exceed one hour. In case of temporal network disconnections, when a HoloLens re-joins the session, it registers as a new client and automatically downloads the shared anchor information to synchronize with the current transformation matrices and states of all the shared objects in the scene.

\subsubsection{Data Inspection}

\begin{figure}[htb]
    \centering
    \includegraphics[width=\linewidth]{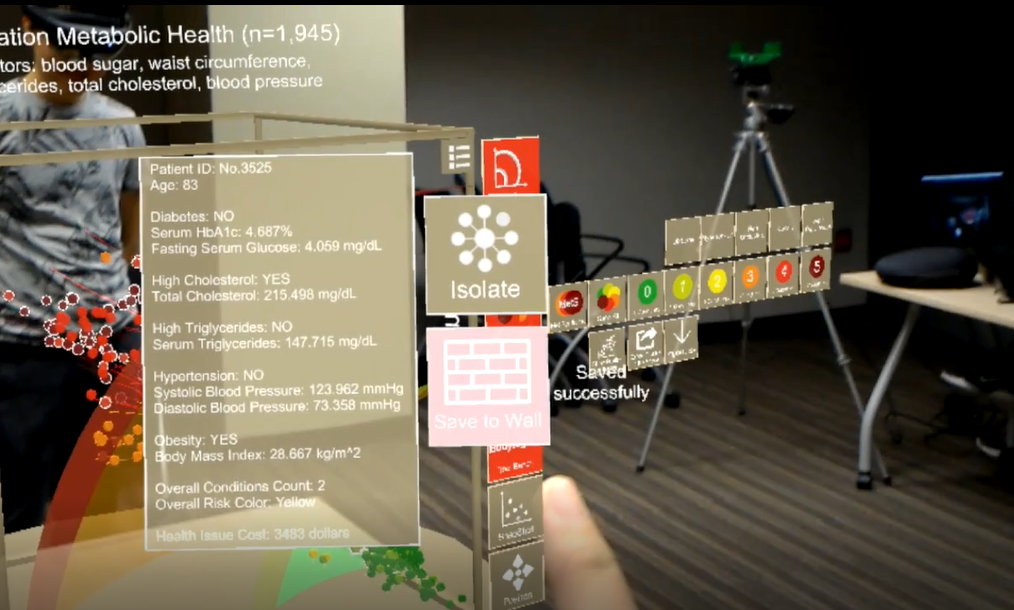}
    \caption{First person view of the HoloLens application with 3D menu panels.}
    \label{fig:datacube-menus}
\end{figure}

Each user can select data points in the data cube by clicking on them. This brings up detailed information about the data point, which represents a specific individual of a population. If a profile that needs further concern is identified, the user can choose to save it to a list that can be exported to the OneDrive folder and studied at a later point. A number of menu items allow changing visualization modes, activating data filters, and many more functions to change how the data is presented to users, as shown in Figure \ref{fig:datacube-menus}.

\subsubsection{Dual-language Support}

To allow non-English speakers to use our DataCube application in Japan, we added dual language support for English and Japanese to our application. We used this feature very successfully in demonstrations to Japanese users in Tokyo. We currently only support two languages, but given a translator we could easily support additional languages.

With the support for multiple languages, when users with different language backgrounds join the same shared experience session, each user can change their own language settings in the menu with the language icon. Since the language configuration is not a shared setting, speakers with different language preferences can see all text in the application in their preferred language. In this way, the app ensures a smooth user experience and facilitates communication when used in an international meeting. 

This is a fascinating feature of AR, which we would love to see more generally used in AR applications, ideally supported on the operating system level. Our multi-language system is implemented with a database which allows each text element in the app to query the corresponding translation when it is created. The database could potentially be adapted to use an online translation service, such as Google Translate to support more languages automatically.

\subsubsection{Google Daydream Controller}

Finger taps in the air can get tiring rather rapidly when using a HoloLens application for a longer duration. The HoloLens comes with a simple clicker, which helps alleviate this issue, but it still requires head orientation to select objects on the screen. We wanted to add an interaction concept which resembled more that of VR headsets with 3D controllers. Therefore, we integrated a Google Daydream \cite{Daydream} controller into our HoloLens application. It allows for pointing in 3D, but the controller does not know its location in 3D space. Our application allows the user to switch switch between hand gestures and the Daydream controller upon request. 

The integration of the Daydream controller extends the user interactions for our application in the following ways. Clicking the button on the Daydream controller is more intuitive and less tiring than the air tap gestures \cite{airtap}. Air taps require a learning curve and have a greater chance of recognition errors than a button click. The orientation tracking of the controller allows for more precise pointing and selection with a virtual laser pointer, which follows the orientation of the controller. This mechanism works much like in actual Google Daydream VR applications.

The mechanism of switching between air taps and the Daydream controller is done automatically. If the user's hand is within the HoloLens's field of view, the app will expect air taps. If the Daydream controller is linked to the HoloLens, and an action is detected from the controller, the app will switch to the Daydream controller. This allows user to seamlessly switch between the two modes without having to use a menu item.

\subsubsection{Spectator View}
\begin{figure}[htb]
    \centering
    \includegraphics[width=\linewidth]{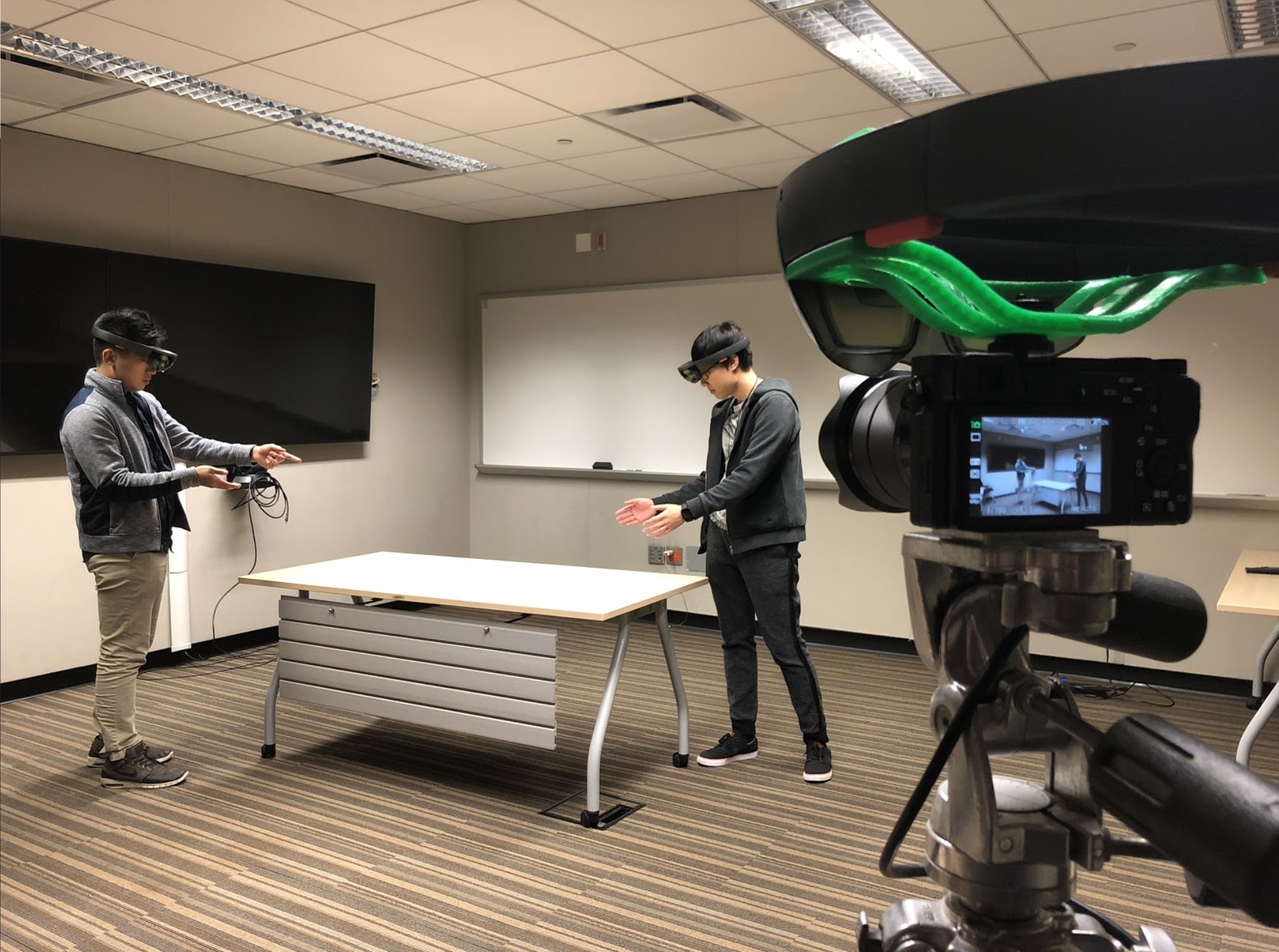}
    \caption{The spectator view camera setup.}
    \label{fig:spectator}
\end{figure}

One typical limitation of augmented reality experiences is the lack of a sense of presence for local observers who are not wearing an AR headset. We help solve this problem by offering a Spectator View \cite{spectator}: a real-time view of a multi-user HoloLens session from a fixed viewpoint, using a Digital Single Lens Reflective camera (DSLR). Figures \ref{fig:datacube} and \ref{fig:datacube-2users} were created with our Spectator View system. In comparison, Figure \ref{fig:datacube-menus} is an image taken directly from the HoloLens system. Figure \ref{fig:spectator} shows the setup of the DSLR camera with the attached HoloLens which we use for capturing the Spectator View. The HoloLens is attached to the flash mount on the top of the camera via a custom 3D printed bracket.

Our spectator setup includes a DSLR camera (Sony alpha-6500), a video capture card (Elgato HD60-S), the 3D printed bracket for mounting the HoloLens onto the camera, and a PC for blending HoloLens and camera video streams. The procedure for our Spectator View is implemented as follows: 

\begin{enumerate}
    \item The PC launches the simulated application via Unity.
    \item The HoloLens transmits real-time position and orientation information to the PC and the simulation app uses it to compute the correct transformation matrices for all virtual objects in the scene.
    \item The DSLR camera transmits high-resolution images through the capture card to the PC and the simulation app blends the camera-captured view with the virtual objects to generate the output image.
    \item We project the output image via a projector to show the real-time AR experience from the camera's perspective to the people in the room who are not wearing HoloLenses.
\end{enumerate}

We choose to use a separate DSLR camera instead of the HoloLens's built-in RGB camera for two reasons. First, we want to present the entire on-going session for the spectators, so we need a camera that has a wider field of view than the HoloLens to capture as much of the room as possible. Second, we need high resolution images to create a video stream of high quality for the projector. In our experience, the spectator view is a very useful feature to allow everyone in a conference setting to follow a data analysis session. 

%\section{Discussion}

\section{Conclusions}

Designing an interactive data visualization tool is always a challenge, especially when trying to make the best use of an AR system. At the time of our research, the Microsoft HoloLens was the best self-contained AR platform available to us. We created a multi-user application for the HoloLens, which can visualize population health data with a 3D data cube, as well as a 2D wall display, designed to be used in a conference room. This application has been successful enough that it has been used in a commercial environment for well over a year. In the future, we hope to port our application to the new HoloLens 2 or other devices with greater fields of view, higher resolution displays, and better hand gesture recognition.

%%%%%%%%%%%%%%%%%%%%%%%%%%%%%%%%%%
% Bibliography
%%%%%%%%%%%%%%%%%%%%%%%%%%%%%%%%%%

%%%%%%%%%%%%%%%%%%%%%%%%%%%%%%%%%%
% Biography
%%%%%%%%%%%%%%%%%%%%%%%%%%%%%%%%%%

\begin{biography}
Wanze (Russell) Xie was an undergraduate student in the computer science department of UCSD when he wrote most of the software for this project. He graduated from UCSD in June of 2019 and is currently a graduate student at Stanford University.

Yining Liang was an undergraduate student at UCSD when the work for this paper was being done, and has since graduated with a B.Sc. degree in computer science.

Janet Johnson is a graduate student in UCSD's department of Computer Science and Engineering and is advised by Dr. Weibel. Her work focuses on human-centered interaction design for augmented reality experiences.

Andrea Mower helped support the Bodylogical and UCSD teams in their experimentation with augmented reality for data visualization and storytelling. She holds a BS from Brigham Young University in Technology and Engineering Education. Her area of specialization is helping companies explore adoption of emerging technology for business use. Other research areas include comparing the effectiveness of training in virtual reality to traditional learning modalities.

Dr. Samuel Burns, Colleen Chelini and Paul D'Alessandro are members of the BodyLogical team.

Dr. Nadir Weibel is an Associate Professor in the Department of Computer Science and Engineering at UC San Diego, and a Research Health Science Specialist in the VA San Diego Health System. His work on Human-Centered Computing is situated at the intersection of computer science, design and the health sciences. He is a computer scientist who investigates tools, techniques and infrastructure supporting the deployment of innovative interactive multi-modal and tangible devices in context, and an ethnographer using novel methods for studying and quantifying the cognitive consequences of the introduction of this technology to everyday life. His research interests range from software engineering to human computer interaction, particularly mobile health, computer supported cooperative work, medical informatics and mobile and ubiquitous computing.

Dr. J\"urgen Schulze is an Associate Research Scientist at UCSD's Qualcomm Institute, and an Associate Adjunct Professor in the computer science department, where he teaches computer graphics and virtual reality. His research interests include applications for virtual and augmented reality systems, 3D human-computer interaction, and medical data visualization. He holds an M.S. degree from the University of Massachusetts and a Ph.D. from the University of Stuttgart (Germany).
\end{biography}


\small
\begin{thebibliography}{11}

\bibitem{Butscher2018}Simon Butscher, Sebastian Hubenschmid, Jens Mueller, Johannes Fuchs, Harald Reiterer; "Clusters, Trends, and Outliers: How Immersive Technologies Can Facilitate the Collaborative Analysis of Multidimensional Data", In Proceedings of ACM CHI 2018, New York, NY, USA, URL:  https://dl.acm.org/citation.cfm?id=3173664
\bibitem{Bodylogical}Bodylogical health simulator, https://www.pwc.com/us/en/industries/ health-industries/library/doublejump/bodylogical-precision.html
\bibitem{HoloLens}R. Furlan, "The future of augmented reality: Hololens - Microsoft’s AR headset shines despite rough edges", Resources\_Tools and Toys, IEEE Spectrum, June 2016.
\bibitem{OnSight}JPL: Mars Virtual Reality Software Wins NASA Award, URL: https://www.jpl.nasa.gov/news/news.php?feature=7249
\bibitem{Millais2018}Patrick Millais, Simon L. Jones, and Ryan Kelly. 2018. "Exploring Data in Virtual Reality: Comparisons with 2D Data Visualizations". In Extended Abstracts of the 2018 CHI Conference on Human Factors in Computing Systems (CHI EA '18). ACM, New York, NY, USA, Paper LBW007, 6 pages. DOI: https://doi.org/10.1145/3170427.3188537 
\bibitem{Daydream}Google Daydream, URL: https://arvr.google.com/daydream/
\bibitem{spectator} Spectator View URL: \\ https://docs.microsoft.com/en-us/windows/mixed-reality/spectator-view
\bibitem{holoLensdoc} Microsoft HoloLens Official Document URL: \\ https://docs.microsoft.com/en-us/windows/mixed-reality/shared-experiences-in-mixed-reality
\bibitem{bach2017drawing} Bach, Benjamin and Sicat, Ronell and Pfister, Hanspeter and Quigley, Aaron, "Drawing into the AR-CANVAS: Designing Embedded Visualizations for Augmented Reality". Workshop on Immersive Analytics, IEEE Vis, 2017
\bibitem{airtap} Markus Funk, Mareike Kritzler, and Florian Michahelles. 2017. HoloLens is more than air Tap: natural and intuitive interaction with holograms. In Proceedings of the Seventh International Conference on the Internet of Things (IoT ’17). Association for Computing Machinery, New York, NY, USA, Article 31, 1–2. DOI:https://doi.org/10.1145/3131542.3140267
\bibitem{Trestioreanu2018}Lucian Trestioreanu, "Holographic Visualisation of Radiology Data and Automated Machine Learning-based Medical Image Segmentation", Master's Thesis at Universite du Luxembourg, August 2018.

\end{thebibliography}
\end{document}